\documentclass[twocolumn]{aastex63}

\def\lapp{\ifmmode\stackrel{<}{_{\sim}}\else$\stackrel{<}{_{\sim}}$\fi}
\def\gapp{\ifmmode\stackrel{>}{_{\sim}}\else$\stackrel{>}{_{\sim}}$\fi}
\usepackage{multirow}
\usepackage{color}
\usepackage{amsmath}
\usepackage{soul}
\usepackage{gensymb}
\usepackage{hyperref}
\usepackage{epstopdf}
\usepackage{lineno}

\submitjournal{ApJ}

\begin{document}

\shorttitle{PSR~J2030+4415's Filament}
\shortauthors{}

\title{The Long Filament of PSR~J2030+4415}

\correspondingauthor{Roger W. Romani}
\email{rwr@astro.stanford.edu}

\author{Martijn de Vries}
\affiliation{Department of Physics/KIPAC, 
Stanford University, Stanford, CA 94305-4060, USA}
\author{Roger W. Romani}
\affiliation{Department of Physics/KIPAC, 
Stanford University, Stanford, CA 94305-4060, USA}

\begin{abstract}
New X-ray and optical observations shed light on the remarkable X-ray filament of the Gamma-ray pulsar PSR~J2030+4415. Images of the associated H$\alpha$ bow shock's evolution over the past decade compared with its velocity structure provide an improved kinematic distance of $\sim$0.5kpc. These velocities also imply that the pulsar spin axis lies $\sim 15^\circ$ from the proper motion axis which is close to the plane of the sky. The multi-bubble shock structure indicates that the bow shock stand-off was compressed to a small value $\sim 20-30$y ago when the pulsar broke through the bow shock to its present bubble. This compression allowed multi-TeV pulsar $e^\pm$ to escape to the external ISM, lighting up an external magnetic field structure as the `filament'. The narrow filament indicates excellent initial confinement and the full $15^\prime$ ($2.2$~pc=7~lt-y) projected length of the filament indicates rapid $e^\pm$ propagation to its end. Spectral variation along the filament suggests that the injected particle energy evolved during the break-through event.
\end{abstract}


\section{Introduction} \label{sec:intro}

In \citet{dr2020} we found that PSR~J2030+4415 (hereafter J2030), a $P=308$\,ms, $\tau = 6\times 10^5$\,yr (${\dot E} = 2.2 \times 10^{34}{\rm\,erg\,s^{-1}}$) radio-quiet $\gamma$-ray pulsar, has a long, narrow X-ray filament. In the discovery image, we see a pulsar wind nebula (PWN) trailed behind the pulsar and a filament at an angle $130^\circ$ to this motion extending $5^\prime$ to the edge of the X-ray image. In this paper we report on a measurement of the filament's full extent as well as supporting optical observations that help probe the nature of the filament's origin.

Only a handful of pulsar X-ray filaments are known. The most famous is associated with the PSR B2224+65/Guitar nebula \citep{Hui2007}. Other large angle long filaments are seen for PSR J1101$-$6101/Lighthouse \citep{Pavan2016} and PSR J1509$-$5830 \cite{Klingler2016} (other candidates have short exposures or are ambiguous). Together with J2030, all four appear to be fast moving pulsars and at least three have H$\alpha$ bow shock structures (the H$\alpha$ image in \citet{Pavan2016} showed too much diffuse H$\alpha$ emission to detect, or exclude, a bow shock). The neutral hydrogen required for the H$\alpha$ emission indicates a high ambient proton density $n_0$, which together with a large pulsar velocity $v$ implies a bow shock with small forward standoff distance $r_0= [{\dot E}/(4\pi \mu m_p n_0 c v^2)]^{1/2}$ for a mean particle mass in proton units $\mu\approx 1.38$. This association with small $r_0$ tends to support the \citet{Bandiera2008} picture for the origin of the Guitar nebula filament: when the standoff distance $r_0$ approaches the gyroradius $r_g = \gamma m_e c^2/eB$ of the relativistic $e^\pm$ generated by the pulsar and its wind termination shock, energetic charges can escape to external magnetic field lines near the apex. Synchrotron emission from particles propagating rapidly along such external fields give rise to the X-ray filaments. These filaments are generally asymmetric, being much longer on one side of the pulsar and its PWN trail. This asymmetry may be associated with the magnetic field direction of the polar flow in the PWN, since reconnection of the forward facing hemisphere tends to favor emission onto field lines of the appropriate polarity. Simulations of particles in magnetosphere termination shocks \citep{Bucciantini2018, Barkov2019, Olmi2019} have illustrated such asymmetry.

\begin{figure*}
\hspace*{-5mm}\includegraphics[scale=0.99]{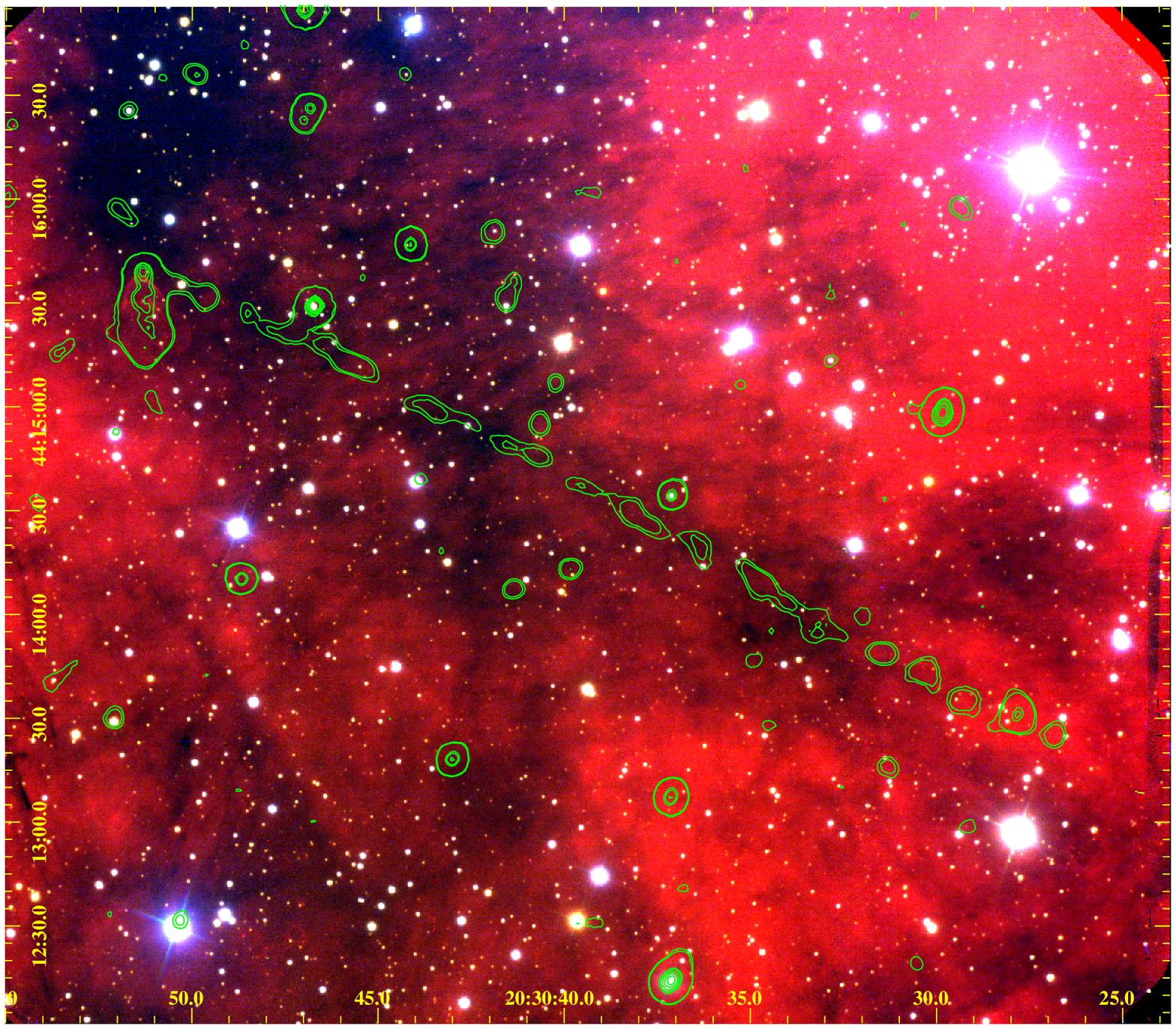}
\vskip -16.67cm\hskip 82mm \includegraphics[scale=0.55]{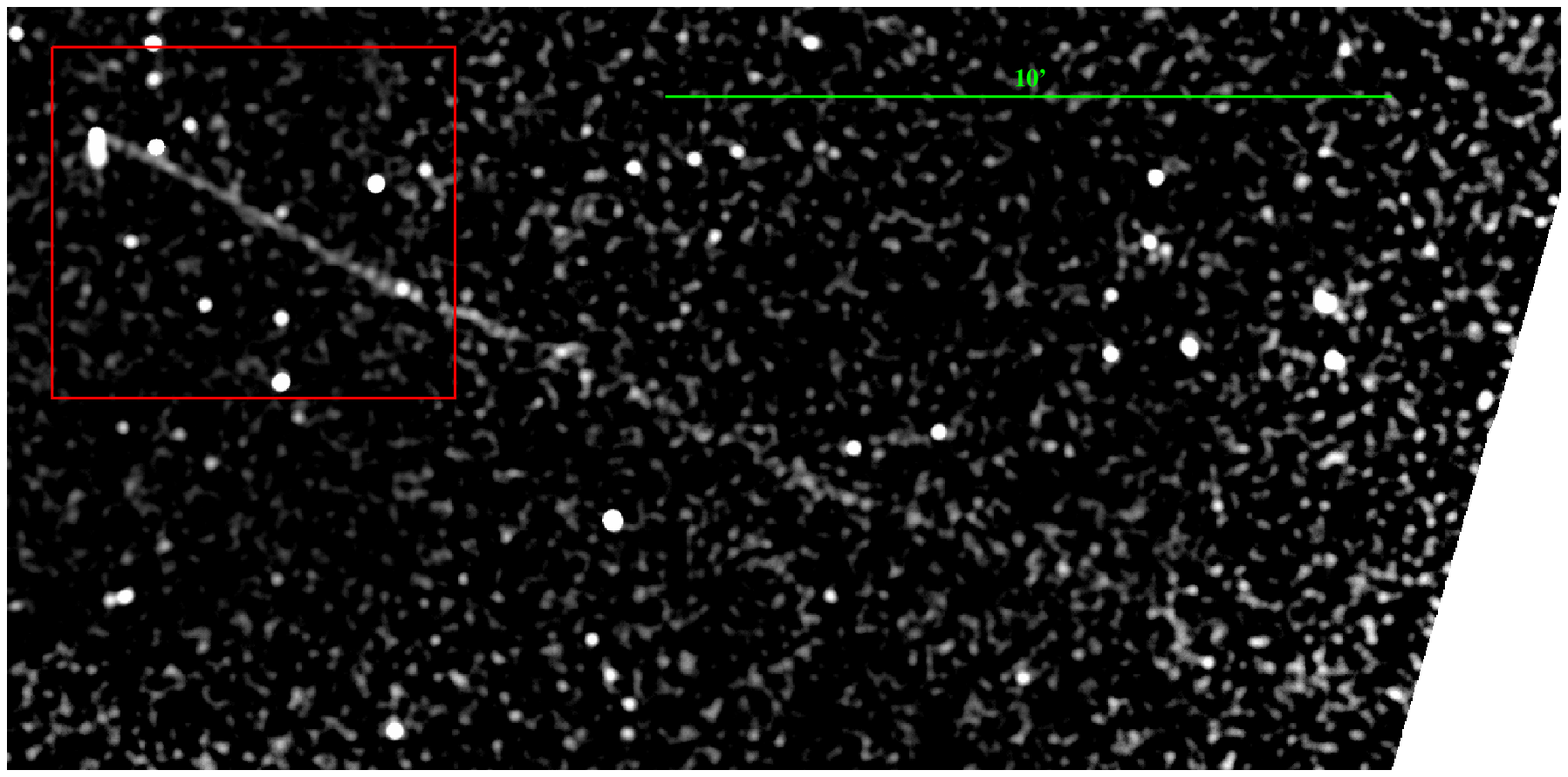}
\vskip 11.5cm
\caption{The PSR J2030+4415 field in GMOS-N H$\alpha$ (red), $r$ (green) and $g$ (blue). The smoothed green contours show 1-5keV CXO ACIS emission; some arises from field stars and background sources, but the PWN and filament are visible. The grey scale inset at upper right shows the combined ACIS data following the filament's 15$^\prime$ length; the red square shows the optical image coverage.}
\label{fig:HaX}
\end{figure*}

J2030's filament provides a good opportunity to explore these ideas: it is long, narrow, seen only on one side and arises from a complex PWN with multiple H$\alpha$ bubbles. Here we describe a new imaging campaign which helps us probe the filament geometry and origin.

\section{Observations} \label{sec:sec2}

\begin{figure*}
\centering
\includegraphics[scale=0.95]{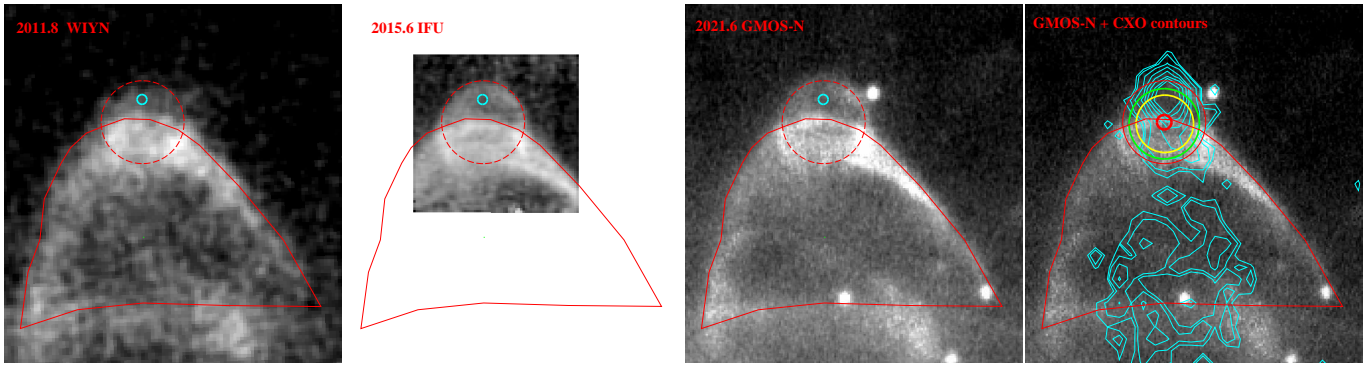}
\caption{\footnotesize The evolution of the apex bubble over a decade. For reference, outlines in each frame mark the forward bow shock (solid red) and the apex bubble (dashed red), as traced from the 2021 data. The panel cover $13.4^{\prime\prime} \times 13.7^{\prime\prime}$ (only $6.8^{\prime\prime} \times 6.5^{\prime\prime}$ for the panel 2 IFU image). The pulsar location is marked with a cyan circle. In the rightmost frame the 2021 image is repeated with estimated bubble radii from the  three epochs shown as yellow, green and red circles. The common center is a red inner circle. Cyan contours show the {\it CXO} flux; the filament is not visible at this fine scale.  }
\label{fig:PWN}
\end{figure*}

\subsection{A new H$\alpha$ bow shock Image}

Gemini/GMOS-N observations of J2030 were taken on August 3 and 9, 2021 (MJD 59429, 59435) under good $0.5-0.6^{\prime\prime}$ imaging conditions (a few usable, but sub-par exposures were also obtained on August 8). The data include $6\times 600$s H$\alpha$ exposures, covering the PWN and the first $5^\prime$ of the filament. Accompanying $3\times 30$s $gri$ images were also obtained for continuum subtraction and characterization of the field stars. Pipeline image reductions proved adequate and we assembled dithered combined images of the field (Figure \ref{fig:HaX}).

It is interesting to compare this latest H$\alpha$ exposure of the nebula apex with the original 2011 WIYN discovery image and the velocity-integrated image from GMOS-N IFU data taken in 2015 (Figure \ref{fig:PWN}). While the bow shock hardly shifts, the `bubble' at the apex of the structure expands noticeably over the past decade. The seeing-limited ground based images (esp.~in 2011) and the limb complexity preclude a direct fit, but matching to the apparent bubble edge gives radii of $\approx 1.18^{\prime\prime}$, $1.45^{\prime\prime}$, and $1.71^{\prime\prime}$ at the three epochs. This gives a mean expansion rate of $\mu_b = 56\pm 8$mas/y, assuming a common center, which is near the tip of the bow shock structure (red circle in the right panel of Figure \ref{fig:PWN}). If the expansion were constant at the last decade's rate it would have initiated from this point in 1989$\pm 1$ (32y ago). Note that DR20 used archival and new {\it CX0} imaging to measure the pulsar proper motion at $85\pm 16$mas/y. It is now $\sim 1.0^{\prime\prime}$ North of the bubble center. Thus the pulsar passed through this point $\sim 12\pm 2$y ago. This suggests that the bubble was initiated by a break-through below this position, somewhat before the 2011 discovery epoch, and that rapid initial bubble expansion out-ran the pulsar, but has now settled down to the present more modest rate. Note however that the bubble expansion rate is only $1.2\sigma$ smaller than the pulsar velocity and that, eventually, we expect that the bubble leading edge motion will settle down to match the pulsar velocity, forming a new stationary bow shock structure. While it seems that we are in transition from rapid bubble expansion at break-out to bow shock formation, additional high resolution imaging will be needed to resolve the bubble kinematics. 

Some kinematic information also comes from the IFU velocity slices (Figure 2 of DR20). The bow shock structure behind the apex bubble is dominated by low velocities. This flux is fairly symmetric over the range $\pm 70$~km/s, indicating that the bow shock axis (and pulsar velocity) lie close to the plane of the sky. Above this, we see the near-circular expanding apex bubble shell, especially in the $\pm$94km/s channels. The bubble extreme velocity peaks are offset N ($+v$, redshifted) and S ($-v$, blue shifted) from the pulsar position. This indicates that the bubble is oblate and expands faster along its major axis. Again S/N, resolution and complexity prevent a direct fit, but matching an oblate spheroid expansion model to the position-velocity cube indicates that the $+v$ major axis points $\approx 10^\circ$ into the plane of the sky at PA$\approx -10^\circ$ (to the west); the model match degrades appreciably for angles $\sim 10^\circ$ from these values. The physical oblateness is $\approx 0.2$, while the velocity oblateness is $\approx 0.5$. In DR20 we attempted to match a \citet{Wilkin2000}-type bow shock model with an equatorial ${\rm cos^2}\theta$ flow; this places the spin axis close to the line-of-sight with the equatorial bulges forming the velocity extrema. Modern pulsar models associate the gamma-ray emission with the outer magnetosphere or the wind zone current sheet, both of which beam to the spin equator. Thus the oblate spheroid model described here, with its equatorial view of the pulsar, seems more satisfactory. We interpret the major axis as marking the polar momentum flux of the pulsar wind. The larger velocity oblateness suggests that the outflow was closer to spherical in the initial outburst and that the pulsar wind latitudinal structure is increasingly affecting the forward shock position. The best-match instantaneous equatorial expansion velocity is $v_e\approx 120$km/s. Comparison with the 2011-2021 expansion proper motion then gives a kinematic pulsar distance of $d=500\,{\rm pc} (\mu_b/56 {\rm mas/y})/(v_e/120 {\rm km/s})$, i.e. about 2/3 of the distance from the heuristic $L_\gamma= (10^{34}{\rm erg\,s^{-1}} {\dot E})^{1/2}$ pulsar luminosity estimate. This makes J2030 one of the closest known energetic PWNe.

\begin{figure*}[t!]
\vskip -0.4truecm
\centering
\includegraphics[width=0.8\linewidth]{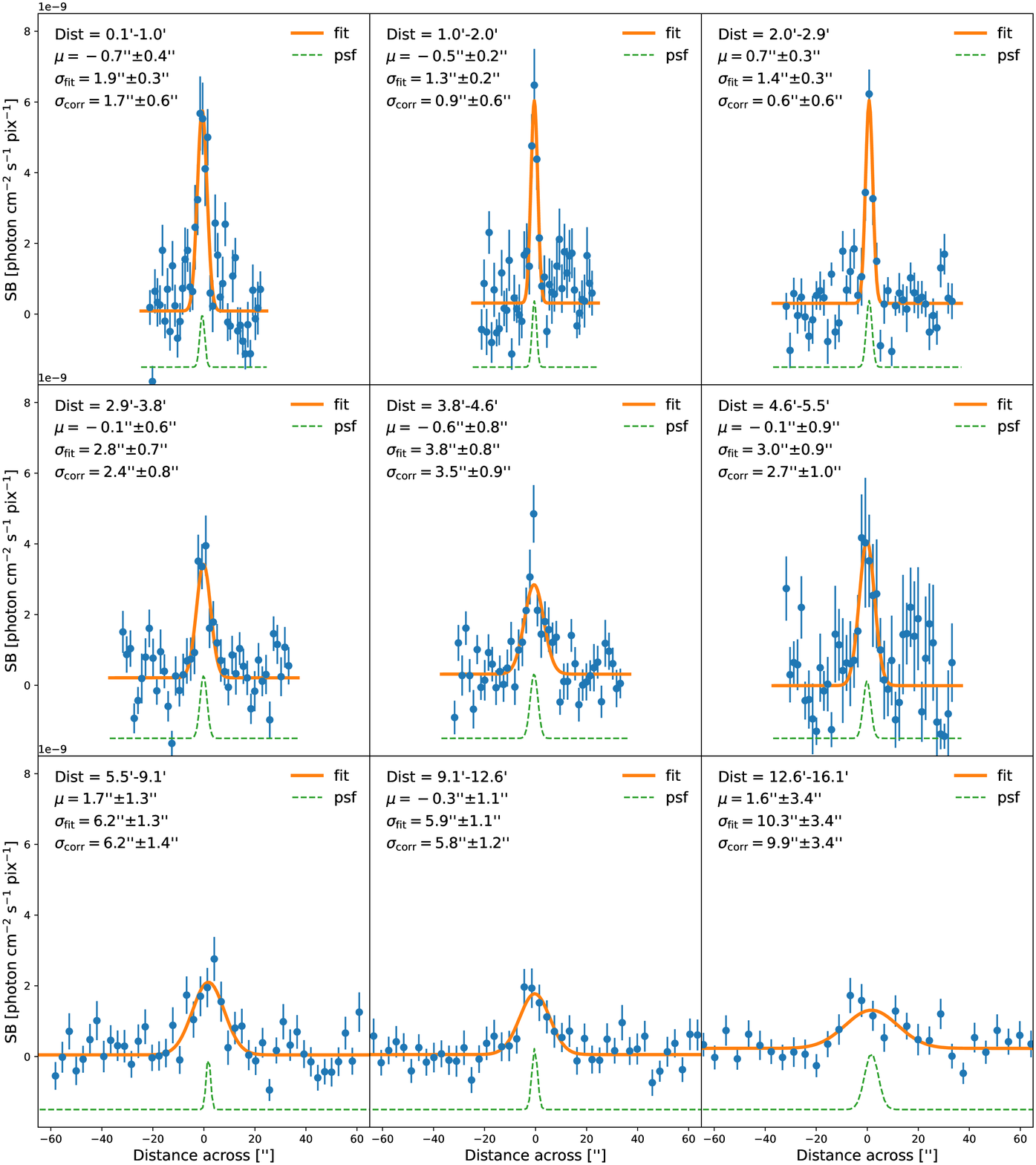}
\caption{\footnotesize Surface brightness profiles across 9 segments of J2030's filament. The x-axis indicates the distance across the filament in arcsec, with the negative direction being behind the filament (south/south-east), and the positive direction ahead of the filament (north/north-west). To each profile, we fit a Gaussian plus a constant background. The green dashed line indicates the average width of the PSF (approximated as a Gaussian) at the location of the segment. Labels on the top left of each panel indicate 1) the distance range of the segment away from the pulsar, 2 and 3) the mean and the standard deviation from the Gaussian fit, and 4) the corrected width, which is obtained by subtracting the PSF width from the fitted width in quadrature and adding a systematic error of $0.5^{\prime \prime}$ to the total uncertainty. }
\label{fig:SBfits}
\end{figure*}

\subsection{New {\it CXO} X-ray Measurements}

In DR20, we reported on four \textit{CXO} exposures from April 10-15, 2019 [ObsIDs 20298 (45\,ks), 22171 (40\,ks), 22172 (45\,ks), 22173 (23\,ks); $\sim 153$\,ks total] and compared with an archival exposure [ObsIDs 14827 (42\,ks)] from April 15, 2014. These placed the pulsar on the S3 chip, and the new filament ran to the edge of the discovery image. We therefore augmented this study with ACIS-I exposures on Feb 13 (44.6ks)  and Nov. 8 (30 ks) 2021 [ObsIDs 23536, 24954 and 24236], with the pointing centered $\sim 5^\prime$ along the filament, allowing us to trace the filament extent to much larger angle. All exposures were in VF mode. The data were processed and analysed using \textit{CIAO} 4.12 and CALDB 4.9.1 and the region of the mosaic covering the filament is shown as the inset in Figure \ref{fig:HaX}. Because the particle background level of the back-illuminated ACIS S-3 chip is significantly higher than the front-illuminated chips in the ACIS array, we created a particle background model using the ACIS `stowed' backgrounds, scaling these backgrounds to the science images using the counts between $9.5-12\,$keV. At this energy range, the effective area of \textit{CXO} is negligible, and the vast majority of counts are particle events. Subtracting this particle background model removes approximately $80$ to $90\%$ of the background flux level and achieves a roughly uniform background level across the field. After background subtraction, we created a merged and exposure-corrected mosaic using the \textit{merge\_obs} tool. For the exposure-corrected image, we used an energy range of $1-5\,$keV to achieve maximum contrast between the filament and the background. Finally, to avoid contamination from point sources in our measurements of the filament surface brightness, we used the CIAO tool \textit{dmfilth} to mask out point sources around the filament and fill them in with Poisson-noise background from a surrounding annulus. 

The pulsar point source and PWN trail are unfortunately far enough off axis that the new exposures do not contribute materially to mapping their complex X-ray structure. A full study would require a very long {\it CXO} exposure (or eventually LynX observations). However the new data do show that the filament extends dramatically further than in the discovery image, fading out 15-16$^\prime$ from the PWN. The filament shows appreciable curvature, with a particularly pronounced southward bend 1$^\prime$ from the pulsar.

In order to measure the properties of the filament, we extracted nine surface brightness profiles across the filament, each perpendicular to a ridge line drawn by eye on the exposure-corrected X-ray image. To each filament segment, we fit a Gaussian plus a constant background, as shown in Figure \ref{fig:SBfits}. We used the CIAO tool \textit{mkpsfmap} to calculate the average PSF width (weighted by the exposure time of each individual observation) at the location of the segment.

Although the filament ridge line was drawn by eye and is thus an imperfect representation of its true location, the fitted Gaussian means are within 1$^{\prime \prime}$ of the estimated ridge line in the inner 6 segments, and even the outer, most diffuse segment does not deviate from the ridge more than 1.7$^{\prime \prime}$. Despite this, a slight systematic error can remain as the result of mismatch between the drawn line and the filament. We therefore add a systematic error of 0.5$^{\prime \prime}$ to the filament width in each region.

\begin{figure*}
\vskip -0.6truecm
\centering
\includegraphics[width=0.99\linewidth]{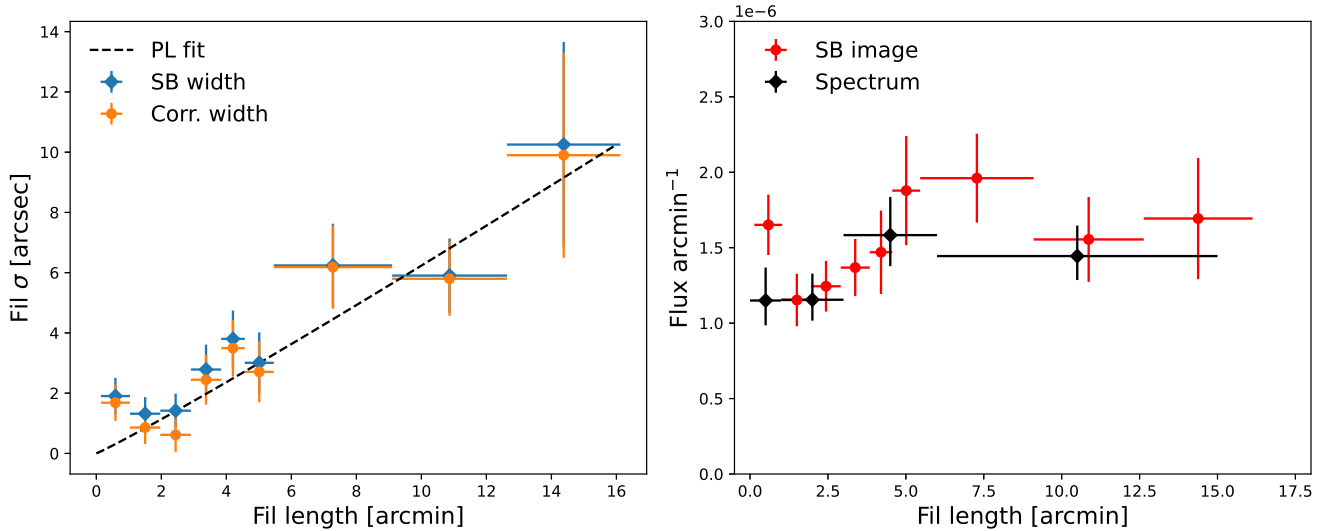}
\caption{\footnotesize \textit{Left panel}: the filament Gaussian width $\sigma$ as a function of distance along the filament. The blue points show the fitted width, while the orange points show the width corrected for PSF size. The dashed black line shows a power-law fit to the orange points, with fitted slope $1.0 \pm 0.2$. \textit{Right panel}: the integrated 1 to 5 keV photon flux per unit length along the filament, in units of $\rm photon\,cm^{-2}\,s^{-1}\,arcmin^{-1}$. The red points indicate the fluxes per arcmin from the surface-brightness analysis, using an exposure map with a $\Gamma=1.5$ energy weighting. The black points indicate the fluxes from the spectral fits listed in Table \ref{table:fits}, converted to 1$-$5 keV photon flux units. }
\label{fig:SBwidth}
\end{figure*}

Figure \ref{fig:SBwidth} shows the filament width and integrated flux per unit length in nine bins along the filament. The corrected widths of each segment were calculated by subtracting the PSF widths in quadrature from the fitted widths. Somewhat surprisingly, we find that the first filament segment is significantly wider than the subsequent two segments - deviating from the general trend of increasing width with increasing distance. This might be due to unrecognized filament curvature in its initial arcmin. Additionally, there could be unaccounted for non-filamentary emission in a halo near the PWN. If we ignore the first segment and fit a power law to the remaining segments, we find a slope $\alpha=1.0 \pm 0.2$. If we include the first point, we find $\alpha=0.9 \pm 0.2$. 

\medskip
\leftline{\bf X-ray Spectrum}
\medskip

With the modest surface brightness and low counts our X-ray spectral results are limited. Nevertheless we can fit simple power law models to various sections of the filament, assuming a constant absorption $N_H = 6 \times 10^{20} {\rm cm^{-2}}$, the value estimated by converting from the $A_g$ of the Bayestar19 3-D dust maps \citep{Green2019}. We extracted spectra from 4 regions along the filament, and combined the extracted spectra from the 5 ACIS-S observations and the extracted spectra from the 3 ACIS-I observations using the CIAO tool \textit{combine\_spectra}. As the background model, we used a local background of spectra extracted from elliptical regions with centers $\approx 1^\prime$ north and south of the filament, which were then subtracted from the filament spectra. The fit results, shown in Table \ref{table:fits}, are consistent with our previous findings in DR20. We detect a softening in the spectrum beyond $3^\prime$, with a significance of around $2 \sigma$. Of course if the true absorption is larger, the $\Gamma$'s will increase (e.g. for $N_H = 2 \times 10^{21} {\rm cm^{-2}}$ all $\Gamma$ increase by $\approx 0.12$), however the spectral index differences persist. The measured fluxes, when converted to units of 1$-$5 keV photon flux agree with the fluxes estimated from the exposure-corrected image (Figure \ref{fig:SBwidth}), with the exception of the region nearest the pulsar. The deviation between the two measurements suggests that perhaps the width in the surface brightness fits is somewhat overestimated. Overall, the $0.5-7\,$keV flux along the filament appears to increase very marginally by about $1\sigma$. 
While it would be interesting to probe spectral variations across the filament, we lack the resolution and sensitivity to do so.

\begin{table}[t!!]
\caption{Spectral fit results for filament sections with an absorbed power law model\textsuperscript{a}}
\centering 
\begin{tabular}{l l l l l} 
\hline\hline 
Region		& Counts &$\Gamma$ &$f_{-15}/$arcmin\textsuperscript{b}& $\chi^2$/DoF \\[0.5ex]  \hline 
Fil (0-15')  &$905 \pm53$ &$1.50 \pm 0.12$  & $8.0_{-0.6}^{+0.7}$ & 175/180 \\
Fil (0-1')	&$122 \pm 15$ &$1.28 \pm 0.29$  & $7.1_{-1.0}^{+1.3}$ & 16/19 \\
Fil (1-3')	&$219 \pm 23$ &$1.09 \pm 0.23$  & $7.5_{-1.0}^{+1.1}$ & 26/30\\
Fil (3-6')	&$294 \pm 32$ &$1.50 \pm 0.22$  & $8.5_{-1.1}^{+1.4}$ & 81/80\\
Fil (6-15')	&$270 \pm 30$ &$1.76 \pm 0.28$  & $8.1_{-0.9}^{+1.1}$ & 45/48 \\
\hline
\end{tabular}
\label{table:fits} 
\leftline{\textsuperscript{a} $N_H$ fixed at $6\times10^{20}{\rm cm^{-2}}$.}
\leftline{\textsuperscript{b} $0.5-7\,$keV unabsorbed fluxes per arcmin filament length}
\leftline{in units of $10^{-15}{\rm erg\,cm^{-2}s^{-1}}\,{\rm arcmin}^{-1}$.}
\vskip -0.5truecm
\end{table}

\section{Filament Morphology and Conclusions}

The best known filaments are those of the Guitar Nebula/PSR B2224+65 \citep{Cordes1993, Bandiera2008} the `Lighthouse'/PSR J1101$-$6101 \citep{Pavan2016} and PSR J1509$-$5058 \citep{Klingler2016}. J2030's filament does have some special features. It is purely one-sided, initially narrow, with an approximate linear expansion, and its flux varies little with distance. These properties are most similar to the filament of PSR J1509$-$5058.   
 
\begin{figure}
\vskip 0.3truecm
\centering
\includegraphics[scale=0.37]{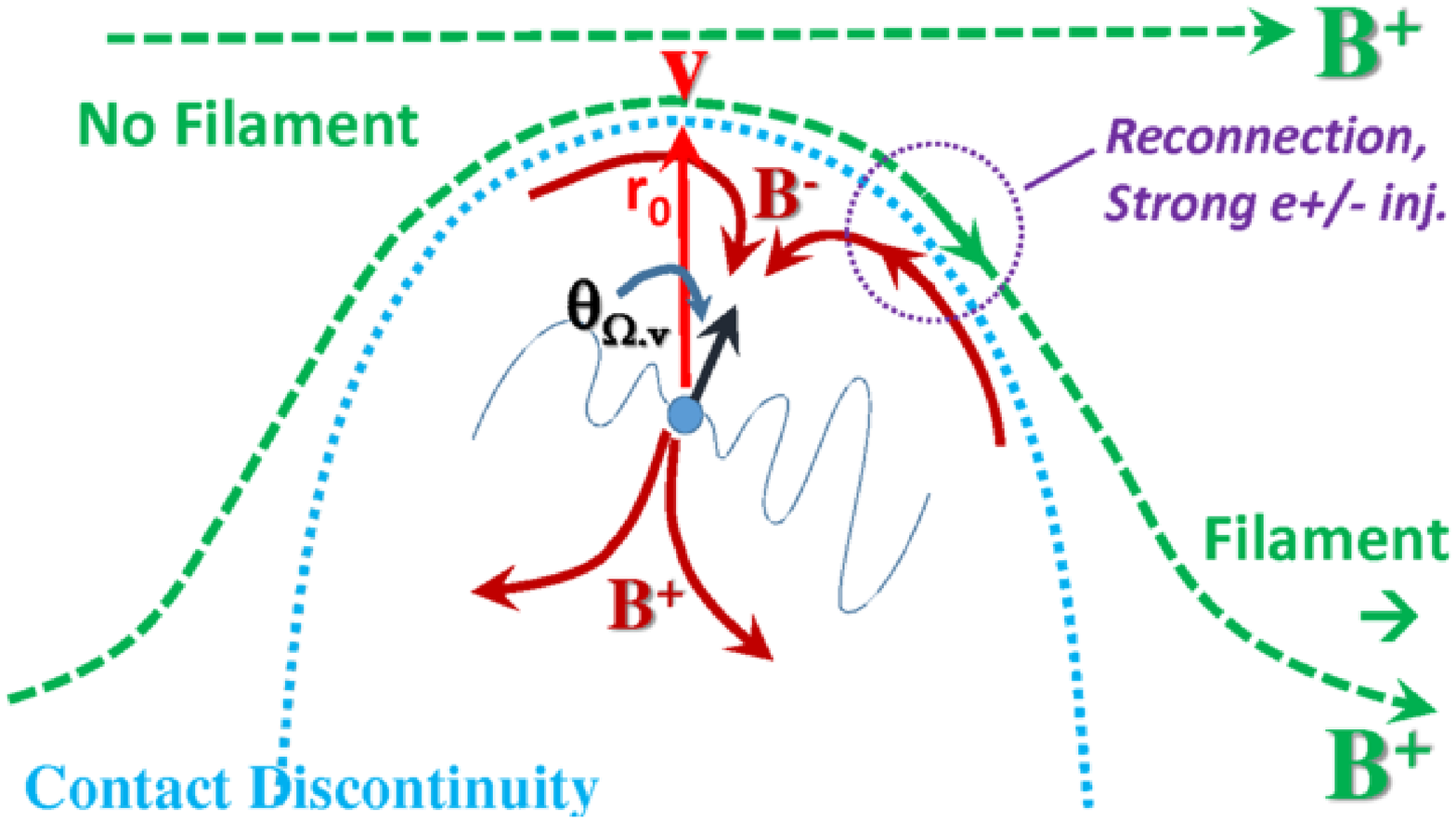}
\caption{\footnotesize Cartoon of magnetic geometry near the magnetopause at the contact discontinuity. When the standoff distance $r_0$ is small, high $\gamma_e$ (large gyroradius) particles can escape to external ISM field lines. This is enhanced by reconnection; when $\theta_{{\vec \Omega} \cdot {\vec v}}$ is $<\pi/2$, reconnection to one side of the bow shock is preferred.
}
\label{fig:cart}
\end{figure}

We would like to connect these properties with the pulsar filament picture outlined in \citet{Bandiera2008} and elaborated by later authors \citep[e.g.][]{Olmi2019, Barkov2019}. J2030's extreme one-sidedness may be connected with the unusually small spin-velocity angle inferred from the H$\alpha$ bubble velocity structure $\theta_{{\vec \Omega} \cdot {\vec v}} \le 15^\circ$; this ensures that one magnetic hemisphere is much closer to the swept-up magnetic field at the apex, fostering reconnection to one sign of ${\vec B}$ (Figure \ref{fig:cart}).

We note that a line following the brightest filament ridge out to $\sim 5^\prime$ passes $\approx 7^{\prime\prime}$ {\it in front} of the pulsar. Yet in the nearest arcmin or so, the filament ridge appears to curve, such that it meets the PSR/PWN axis several arcsec {\it behind} the pulsar (as can be seen in Figure \ref{fig:HaX}). This is close to the bow shock apex and the bubble expansion point. A plausible interpretation is that filament follows a field line that connected with the bow shock apex several decades ago, when $r_0$ was minimal just before break-out to the new forward bubble structure. The pulsar-trailing connection position might alternatively be associated with field line `draping' as the magnetic lines are pushed forward by the bow shock as seen in e.g. the lighthouse nebula \citep{Klingler2016}. However, draping would be expected to produce field line curvature opposite that seen for J2030, so perhaps the curvature is intrinsic to the external field.

The narrow initial filament width places some limits on the guiding magnetic field, since a gyroradius $\rho = 1.6 \times 10^{17} \gamma_8/B_{\mu G}$\,cm subtends $10.7^{\prime\prime} \gamma_8/(B_{\mu G}d_{\rm kpc})$, for an $e^\pm$ energy $10^8\gamma_8 m_e c^2$ and magnetic field $10^{-6} B_{\mu G}$\,G at a distance $d_{\rm kpc}$kpc. As usual, using the synchrotron peak energy $E_{\rm keV}$keV, we eliminate $\gamma_8 = 2.4 (E_{\rm keV}B_{\mu G})^{1/2}$ to write $\theta_{\rm gyro}=26^{\prime\prime}E_{\rm keV}^{1/2}B_{\mu G}^{-3/2}d_{\rm kpc}^{-1}$. Thus using the minimum corrected width and a mean photon energy of 2\,keV, we infer that the filament has $B>20\mu$G at our estimated $d_{\rm kpc}\approx 0.5$. Imperfect ridge line estimates increases this lower bound. Of course we can also estimate the filament field using our spectral measurements and the usual equipartition synchrotron sums. The observed flux from a cylindrical emission region of radius $\theta_w$ and length $\theta_l$ corresponds to a volume emissivity $J=4\pi f d^2/V = 1.9 \times 10^{-22} f_{-15} /(\theta_w^2\theta_l d_{\rm kpc})$\,erg\,s$^{-1}$\,cm$^{-3}$ (for $\theta_{w}$ in arcsec, $\theta_{l}$ in arcmin) measured over $E_1 - E_2$ from a full spectral range $E_{\rm min} - E_{\rm max}$. Assuming a magnetization $\sigma=w_B/w_e$ and measuring $J_{-20}$ in units of $10^{-20}$\,erg\,s$^{-1}$\,cm$^{-3}$ we obtain 
\begin{equation}
\label{eq:syncB}
B = 46 \left[ J_{\rm -20}(E_1,E_2) \sigma \frac{C_{1.5-\Gamma}(E_m, E_M)} {C_{2-\Gamma}(E_1,E_2)}\right] ^{2/7} \mu G
\end{equation}
where $C_q(x_1,x_2) = \frac{x_2^{q} - x_1^q}{q}$, $\Gamma$ is the spectral index and energies are in keV. Using $\Gamma\approx 1.5$ over an assumed full range of 0.01-10\,keV, $f_{-15}\sim 8$/arcmin from Table \ref{table:fits} and a typical filament radius of $\theta_w=2^{\prime\prime}$, we obtain $B\approx 26(w_B/w_e)^{2/7}\mu$G, in reasonable agreement with the minimum width constraint.

In Table \ref{table:fits}, the nearest two zones appear to be harder by $\Delta \Gamma\approx 0.5$ than the farther zones. Although the difference is marginally ($\sim 2 \sigma$) significant, it is tempting to interpret this as a cooling break. Similarly, in DR20 we found weak evidence for a $\tau=1.52\pm0.59^{\prime\prime}$ exponential tail behind the relatively sharp leading edge, again suggesting cooling. However, with an observed proper motion of 85\,mas/y, this exponential tail would correspond to a cooling time of $\sim 18$\,y, which for $\tau_c = 8\times 10^4{\rm y}\, E_{\rm keV}^{-1/2} B_{\mu G}^{-3/2}$ requires a field of $> 200\mu$\,G. Such a large field is not typical of the ISM. 

Moreover, in Table \ref{table:fits} and Figure \ref{fig:SBfits} the filament does not fade along its length as might be expected for cooling. Indeed, the flux/arcmin increases, if anything, in the outer segment. The brightness profile also argues against simple diffusion from on-going injection of pulsar particles filling the filament since, again, one would expect a flux decrease with distance (age). Instead we adopt the more typical $\sim 20-30\mu$G field from the gyroradius and spectral estimates (still a factor of a few larger than general ISM fields). We can then interpret the narrow initial filament width as indicating a short $1^{\prime\prime}/85$\,mas/y$\leq 12$y period of particle injection, during which the pulsar traveled less than 1 arcsec. This injection is likely associated with the bow shock break-through. Thus, the X-ray producing filament particles have been resident for the 20-30y since the breakthrough event, with adequate time to travel the $\approx 15^\prime \approx 7$\,lt-y (at $d_{\rm kpc}=0.5$) to the filament end at $v<c/3$. Little cooling has occurred, since $\tau_c \approx 350$\,y for $B=30\mu$G. 

\begin{figure}
\centering
\includegraphics[width=0.99\linewidth]{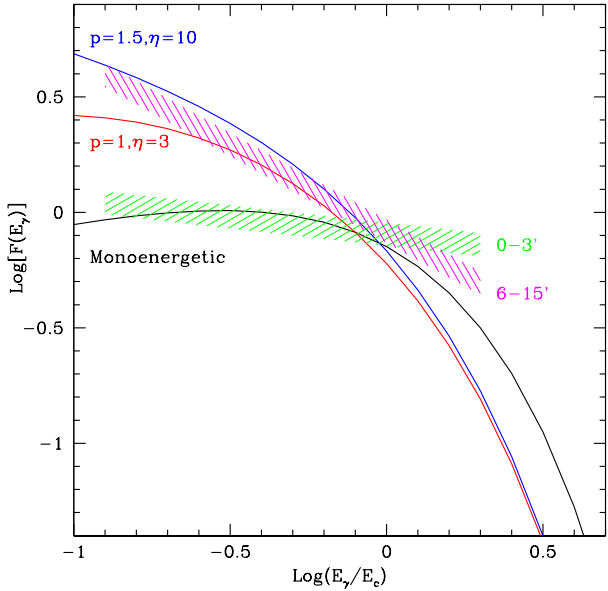}
\caption{\footnotesize Variation in the X-ray spectral index near the synchrotron cutoff (here $\gamma_{e,{\rm max}}$ is set so $E_c=4$\,keV), with varying pulsar injection. The PL fit to the near-pulsar filament 0.5-8\,keV spectrum (green shaded region) may be produced near the synchrotron peak of a mono-energetic particle distribution. The softer emission from more distant (older) zones (magenta shaded region) can be produced by a small range of particles $\gamma_{e,{\rm max}}/\gamma_{e,{\rm min}}=\eta$ with power-law index $p$. Relatively hard $p\approx 1-1.5$ are needed to produce the observed small X-ray $\Gamma \approx 1.2-1.8$. }
\label{fig:inj}
\end{figure}

Still, the electrons near the pulsar should be somewhat younger than those at the far end of the filament. In this picture the (tentative) evidence for $\Gamma$ and flux increase along the filament might be ascribed to variation in the injection process. Let us assume that the pulsar/wind termination shock produces a spectrum of energetic $e^\pm$ $dN_e/d\gamma_e = C \gamma_e^{-p}$ up to some $\gamma_{e,{\rm max}}$. When the standoff distance $r_0$ was compressed to its minimum value at the breakthrough event (where the external density is maximal in the bow shock-compressed ISM) a modest range $\eta$ of the largest energies $> \gamma_{e,\,{\rm min}} = \gamma_{e,{\rm max}}/\eta$ would have sufficiently large gyroradii to escape to the filament. As the apex bubble grew, with the pulsar wind termination shock encroaching on the lower density general ISM, $r_0$ increased and the escaping $\gamma_e$ range (and particle flux) decreased until only $\gamma_{e,{\rm max}}$ could escape to the filament. Continued increase of $r_0$ would shut off all particle injection. Thus the last particles injected (nearest the pulsar) would have a nearly mono-energetic $\gamma_{\rm max}$ spectrum, while more distant filament zones would be dominated by a larger particle population with $\eta$ of a few. Figure \ref{fig:inj} shows how this can produce a spectrum that softens with distance along the filament. The higher number of particles injected at break-through can compensate for diffusion spreading, keeping the distant zones of the filament bright.

We can similarly understand the approximately linear increase of $\theta_w$ along the filament, if particles further along the filament are older, reaching their position by flow and/or diffusion. One could attribute the linear spreading of Figure \ref{fig:SBwidth} to cross-field diffusion, with the small $\theta_w/\theta_l \approx 0.01 \approx (D_\perp/D_\parallel)^{1/2} \approx M_A^2$ attributed to a small cross-field diffusion coefficient $D_\perp$. This suggests a relatively small magnetic Alfven number $M_A\approx 0.1$. Again, stronger injection at break-through might explain the lack of diffusion-induced surface-brightness fall-off, although fine tuning might be needed to match the profile of Figure \ref{fig:SBwidth}.
\bigskip

Our revised picture of the J2030 bow shock and filament suggest rapid evolution of these structures on a decade time scale and sub-arcsec angular scale. This makes precision measurements of the shock structure and filament connection very difficult with ground based seeing. High resolution H$\alpha$ images with {\it HST} could help, but since the break-through event was a few decades in the past such data can primarily constrain the current shock evolution. We still have a challenging task to connect the filament structure with the dynamics of new bubble formation. One important clue could come from a future high S/N X-ray filament measurement. At 85mas/y the pulsar displacement is substantial on the decade timescale and if filament injection is continuous rather than single-epoch, we would expect illumination of new field lines ahead of the present edge. This is indeed what we see for the Guitar nebula  \citep[e.g.][de Vries et al., in prep]{Wang2021} and is feasibly constrained with a deep future {\it CXO} exposure. A lack of such shifted emission would support the break-through injection picture proposed here. Even more valuable would be a high S/N spatially resolved spectral study of the filament to measure synchrotron spectral variations along and across its width. Figure \ref{fig:inj} implies that, if we are observing near $E_c$, there should be substantial spectral curvature across the soft X-ray band. Measurement of such curvature requires substantially higher S/N; the long-field spectra and their variation might be probed with {\it XMM} but probe of spectral variation across the filament width will probably need LynX-era sensitivities. 

Perhaps the most useful prediction of our study is that other H$\alpha$ pulsar bow shocks with small $r_0$ should experience short periods of enhanced injection, causing filaments to appear or brighten. Monitoring these rare bow shocks, e.g. with H$\alpha$ imaging, could alert us to incipient events, triggering extensive X-ray imaging campaigns to study the injected particles and their evolution. If, in addition, we can use pulsar polarization, PWN jet structures and H$\alpha$-revealed PWN momentum asymmetries (as for J2030) to infer 3D geometries and spin-velocity angles, we will then have a powerful tool for probing the conditions needed for efficient reconnection and multi-TeV $e^\pm$ escape. This will be a useful guide for RMHD shock modeler and may have important implications for the propagation of pulsar cosmic ray positrons through the nearby ISM to Earth detectors.

\acknowledgments
We wish to thank the staffs of the observatories who helped in planning for the exposures described in this paper, especially Jean Connelly at the CfA for help with {\it CXO} scheduling and Siyi Xu for assistance with the Gemini observations.

This work was supported in part by NASA grants G08-19049X and G01-22054X administered by the SAO.

\vspace{5mm}
\facilities{CXO, Gemini - GMOS}

\bigskip
\bibliographystyle{aasjournal}
\bibliography{J2030}

\end{document}